\documentclass{jpp}

\usepackage{color}
\usepackage{graphicx}
\usepackage{natbib}

\ifCUPmtlplainloaded \else
  \checkfont{eurm10}
  \iffontfound
    \IfFileExists{upmath.sty}
      {\typeout{^^JFound AMS Euler Roman fonts on the system,
                   using the 'upmath' package.^^J}%
       \usepackage{upmath}}
      {\typeout{^^JFound AMS Euler Roman fonts on the system, but you
                   dont seem to have the}%
       \typeout{'upmath' package installed. JPP.cls can take advantage
                 of these fonts, if you use 'upmath' package.^^J}%
      }
  \else
  \fi
\fi

\ifCUPmtlplainloaded \else
  \checkfont{msam10}
  \iffontfound
    \IfFileExists{amssymb.sty}
      {\typeout{^^JFound AMS Symbol fonts on the system, using the
                'amssymb' package.^^J}%
       \usepackage{amssymb}%
         \let\leq=\leqslant
         
      }{}
  \fi
\fi

\ifCUPmtlplainloaded \else
  \IfFileExists{amsbsy.sty}
    {\typeout{^^JFound the 'amsbsy' package on the system, using it.^^J}%
     \usepackage{amsbsy}}
    {}
\fi

\newsavebox{\astrutbox}
\sbox{\astrutbox}{\rule[-5pt]{0pt}{20pt}}

\title[Density jump as a function of the field for parallel relativistic collisionless shocks]{Density jump as a function of the field for parallel relativistic collisionless shocks}

\author[A. Bret and R. Narayan]%
{Antoine Bret$^{1,2}$, Ramesh Narayan$^{3,4}$%
  \thanks{Email address for correspondence: antoineclaude.bret@uclm.es}
}

\affiliation{$^1$ETSI Industriales, Universidad de Castilla-La Mancha, 13071 Ciudad Real, Spain\\[\affilskip]
$^2$Instituto de Investigaciones Energ\'{e}ticas y Aplicaciones Industriales, Campus Universitario de Ciudad Real, 13071 Ciudad Real, Spain\\[\affilskip]
$^3$ Center for Astrophysics | Harvard \& Smithsonian, Harvard University, 60 Garden St., Cambridge, MA 02138, USA\\
$^4$Black Hole Initiative at Harvard University, 20 Garden St., Cambridge, MA 02138, USA
}

\date{?; revised ?; accepted ?. - To be entered by editorial office}

\begin{document}

\maketitle

\begin{abstract}
Collisionless shocks are frequently analyzed using the magnetohydrodynamic formalism (MHD), even though the required collisionality hypothesis is not fulfilled. In a previous work \citep{BretJPP2018}, we presented a model of collisionless shock displaying an important departure from the expected MHD behavior, in the case of a strong flow aligned magnetic field. This model was non-relativistic. Here, it is extended to the relativistic regime, considering zero upstream pressure and upstream Lorentz factor $\gg 1$. The result agrees satisfactorily with Particle-in-Cell simulations and shows a similar, and important, departure from the MHD prediction. In the strong field regime, the density jump $r$, seen in the downstream frame, behaves like $r \sim 2 + 1/\gamma_{\mathrm{up}}$  while MHD predicts 4 ($\gamma_{\mathrm{up}}$ is the Lorentz factor of the upstream measured in the downstream frame). Only pair plasmas are considered.
\end{abstract}

\maketitle

\section{Introduction}
Shock waves are fundamental phenomena in fluids. In an electrically neutral medium, only binary collisions can mediate the transition between the upstream and the downstream. Consequently, the width of the shock front is of the order of a few mean free paths \citep{Zeldovich}. In a plasma where the mean free path is greater than the dimensions of the system, shock waves can also propagate, mediated by collective electromagnetic effects \citep{Sagdeev66}. Such shock waves are called ``collisionless shocks’’.

When it comes to analysing such shocks, the jump equations of magnetohydrodynamics (MHD) are frequently used, even though this formalism assumes a small mean path since it  ultimately relies on fluid mechanics. It is therefore important to know to which extent the use of MHD is legitimate, or not, when it comes to analysing collisionless shocks.

In \cite{BretJPP2018}, we developed a model capable of predicting the density jump of a collisionless shock in the presence of a parallel magnetic field. The interest of this setup is that according to MHD, the magnetic field should play no role in the density jump of a parallel shock \citep{Kulsrud2005}. However, our model showed that for strong enough a field, the density jump of a strong sonic shock can go from 4 to 2. This theory has been validated by Particle-in-Cell (PIC) simulations in \cite{Haggerty2022}.

The model of \cite{BretJPP2018} was non-relativistic. However, even in the relativistic case, a sharp reduction in the density jump, compared with MHD predictions, was also observed in \cite{BretJPP2017}.

The aim of the present work is to develop the relativistic version of the theory described in \cite{BretJPP2018}, and compare the result with the PIC simulations of  \cite{BretJPP2017}.

In the non-relativistic regime, in line with Buckingham's $\pi$ theorem \citep{Buckingham}, the number of parameters and basic units allows  to compute the density jump in terms of 2 dimensionless parameters only: the upstream sonic Mach number $M_s$ and the magnetic parameters $\sigma$, defined below by Eq. (\ref{eq:sigma})\footnote{5 parameters ($\rho_1,v_1,P_1,B_0,q$), where $q$ is the elementary charge. 3 basic units ($m, kg, s$), hence, $5-3=2$ dimensionless parameters governing the problem.}. In the present relativistic regime, the speed of light is added to the list of parameters. Conversely, the problem must be solved in terms of 2+1 dimensionless parameters: $M_s$, $\sigma$ and $\gamma_1$, the upstream Lorentz factor. In the sequel, we shall only consider the strong sonic shock case, namely $M_s=\infty$, having only $\sigma$ and $\gamma_1$ left as free parameters.

Since the non-relativistic limit has been dealt with in \cite{BretJPP2018}, we shall here only explore the regime $\gamma_1 \gg 1$.

Sections \ref{sec:method} and \ref{sec:conser} summarize the physics of the theory and the relativistic conservation equations needed. Then the problem is solved from Sections \ref{sec:S1} to \ref{sec:S1S2}, in the front frame. Yet, because the PIC simulations were performed in the downstream frame, our solution must be boosted to this frame. This is achieved in Section \ref{sec:PIC}.

\begin{figure}
\begin{center}
 \includegraphics[width=\textwidth]{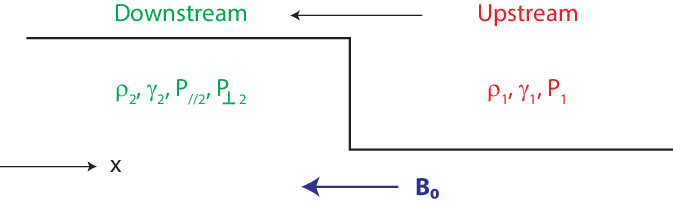}
\end{center}
\caption{The densities $\rho_i$ and pressures $P_i$ are measured in the fluids (upstream and downstream) rest frame. Lorenz factors $\gamma_i$ are measured in the front frame. The upstream is assumed isotropic, but not the downstream. Only the case of a strong shock, namely $P_1=0$ (sonic Mach number $M_s=\infty$), and $\gamma_1 \gg 1$, is studied in this work.}\label{fig:setup}
\end{figure}

\section{Method}\label{sec:method}
For the present work to be self-contained, the method described in \cite{BretJPP2018} is here reminded.

The basic idea is that as it goes through the front, the upstream plasma undergoes an anisotropic compression: it is mainly compressed in the direction parallel to the motion. The resulting downstream is therefore anisotropic, with an increased temperature perpendicular to the motion and the field, while the perpendicular temperature has been conserved.

This stage of the downstream is called ``Stage 1’’.  In a collisional fluid, binary collisions would quickly restore isotropy on a time scale given by the collision frequency. In a non-magnetized collisionless medium, the Weibel instability would also restore isotropy, on a time scale given by the growth rate \citep{Weibel,SilvaPRE2021}. But in the presence of an ambient magnetic field, both theory and solar wind observations show that an anisotropy can definitely be stable \citep{Gary1993,BalePRL2009,MarucaPRL2011}. This is the very reason why the behavior of a collisionless shock can differ from its fluid counterpart. Note that shock accelerated particles, neglected here, can also trigger a departure from MHD \citep{HaggertyApJ2020,BretApJ2020}.

If the field is strong enough to stabilize it, then Stage 1 is the end state of the downstream. On the contrary, the plasma migrates to the instability threshold\footnote{Firehose, see Section \ref{sec:stab}.}, namely, by definition ``Stage 2’’.

Whether we characterize Stage 1 or 2, the density jump is computed from the \emph{anisotropic} MHD equations \citep{Hudson1970}. Yet, within these equations, the anisotropy degree of the downstream is left as a free parameter. The process described above specifically  aims at determining this parameter. More precisely, there are 3 MHD jump equations for the parallel case. But the downstream state involves 4 unknowns\footnote{See Section \ref{sec:conser}.}. Therefore, an additional equation is needed to solve the problem. For Stage 1, this missing equation is simply $T_{\perp 2}=T_{\perp 1}$. For Stage 2, the marginal stability requirement provides the missing equation.

The formalism was developed for a pair plasma. It allows to consider only one single parallel temperature and one single perpendicular temperature instead of 4 different temperature parameters for ions and electrons, as species of different mass can be heated differently at the front crossing \citep{Guo2017,Guo2018}. Yet, our strategy eventually relies mainly on the macroscopic MHD formalism. This allows to think that it could be extended unchanged to electron/ion plasmas, only re-scaling some dimensionless parameters. PIC simulations are currently performed to confirm this point \citep{Shalaby2025}. Yet, in the present work, we shall stick  to pair plasmas.

\section{Conservation equations}\label{sec:conser}
For a parallel shock\footnote{See \cite{Kennel1984} for a perpendicular shock.}, the relativistic conservation equations in the front frame have been derived in \cite{Double2004,Gerbig2011}. They read,
\begin{eqnarray}\label{eq:conser}
  \left[\rho \gamma\beta\right] &=& 0,\\
  \left[\gamma^2\beta(e+P_\parallel)\right]  &=& 0, \nonumber\\
  \left[\gamma^2\beta^2(e+P_\parallel) +P_\parallel \right]  &=&  0, \nonumber\\
  \left[ B  \right]  &=&  0, \nonumber
\end{eqnarray}
where $e, \rho, P$ are the internal energy, the mass density and the pressure of the upstream and downstream fluids in their rest frame. $\gamma$ is the Lorentz factor of the fluid in the front frame, and $\beta=\sqrt{1+\gamma^{-2}}$. The last equation tells the field is conserved at the interface, as is the case for a non-relativistic system \citep{Majorana1987,Kulsrud2005}. We shall therefore drop it when writing these equations in the sequel. In addition, these equations imply that for such an orientation of the field, the MHD problem is eventually equivalent to the fluid problem.

The internal energy $e$  for an anisotropic relativistic gas reads \citep{Double2004,Gerbig2011},
\begin{equation}\label{eq:e_ani}
e = \frac{2P_\perp + P_\parallel}{3(\Gamma - 1)} +  \rho c^2.
\end{equation}

Strictly speaking, the adiabatic index $\Gamma$ is a function of the upstream quantities since depending on them, the downstream can be relativistic in its rest frame, or not\footnote{With $P_1=0$, the adiabatic index of the upstream is not an issue.}. A non-relativistic downstream has $\Gamma= 5/3$, and $\Gamma= 4/3$ if it is relativistic. Because we here consider $\gamma_1 \gg 1$, we shall consider $\Gamma=4/3$ in the sequel.

The first 3 equations contains 4 unknowns: $\rho_2$, $\gamma_2$, $P_{\parallel 2}$ and $P_{\perp 2}$. The Stage 1 \& 2 constraints on $P_{\perp,\parallel 2}$ then provide the fourth equation allowing to solve the problem.

The problem will be dealt with in terms of the following dimensionless variables,
\begin{eqnarray}
 r &=& \frac{\rho_2}{\rho_1}, \label{eq:r} \\
 \theta_{\perp,\parallel} &=& \frac{P_{\perp,\parallel}}{\rho_1c^2},\label{eq:theta} \\
 \sigma &=& \frac{B_0^2/4\pi}{\gamma_1\rho_1c^2}, \label{eq:sigma} \\
 A_2 &=& \frac{P_{\perp 2}}{P_{\parallel 2}}, \label{eq:A2}
\end{eqnarray}
which stand for the density ratio, the dimensionless pressures, the magnetic field parameter\footnote{Although the field does not appear in the conservation equations, it does play a role in the analysis of Stage 2. See Section \ref{sec:S2}.} and the downstream anisotropy respectively.

Accounting for $P_1=0$, Eqs. (\ref{eq:conser}) read,
\begin{eqnarray}
\rho_1 \gamma_1\beta_1 &=& \rho_2 \gamma_2\beta_2,  \label{eq:conserv1} \\
  \gamma_1^2\beta_1  &=&
   \gamma_2^2\beta_2\left(\frac{2\theta_{\perp 2} + \theta_{\parallel 2}}{3(\Gamma - 1)} +   \frac{\rho_2}{\rho_1}  +\theta_{\parallel 2}\right), \label{eq:conserv2}\\
  \gamma_1^2\beta_1^2   &=& \gamma_2^2\beta_2^2\left(\frac{2\theta_{\perp 2} + \theta_{\parallel 2}}{3(\Gamma - 1)} +   \frac{\rho_2}{\rho_1}  +\theta_{\parallel 2}\right) +\theta_{\parallel 2}. \label{eq:conserv3}
\end{eqnarray}
Using the first one to derive,
\begin{equation}\label{eq:rbase}
r = \frac{\rho_2}{\rho_1} = \frac{\gamma_1\beta_1}{\gamma_2\beta_2},
\end{equation}
and using it to eliminate $\rho_2$ in the second and the third, we obtain the system we shall solve for both Stage 1 and Stage 2,
\begin{eqnarray}
  \gamma_1^2\beta_1  &=&
   \gamma_2^2\beta_2\left(\frac{2\theta_{\perp 2} + \theta_{\parallel 2}}{3(\Gamma - 1)} +    \frac{\gamma_1\beta_1}{\gamma_2\beta_2}  +\theta_{\parallel 2}\right), \label{eq:conser2} \\
  \gamma_1^2\beta_1^2   &=& \gamma_2^2\beta_2^2\left(\frac{2\theta_{\perp 2} + \theta_{\parallel 2}}{3(\Gamma - 1)} +   \frac{\gamma_1\beta_1}{\gamma_2\beta_2}  +\theta_{\parallel 2}\right) +\theta_{\parallel 2}. \label{eq:conser3}
\end{eqnarray}

\section{Stage 1}\label{sec:S1}
Here we assume $T_{\perp}$ is conserved, so that $T_{\perp 2}=T_{\perp 1}=0$. We therefore set $\theta_{\perp 2}=0$ in Eqs. (\ref{eq:conser2},\ref{eq:conser3}).
Eliminating $\theta_{\parallel 2}$ between (\ref{eq:conser2}) and (\ref{eq:conser3}) gives the equation for $\beta_2$,
\begin{equation}
(3 \Gamma -2) \gamma_1 \gamma_2^2 \beta_2 (\beta_2-\beta_1)+3 (\Gamma -1) (\gamma_1- \gamma_2)=0.
\end{equation}
Considering in addition $\beta_1 \sim 1$ and $\Gamma=4/3$, it simplifies to,
\begin{equation}\label{eq:beta2S1}
\sqrt{1-\beta_2^2}-(\beta_2-1)^2 \gamma_1 = 0.
\end{equation}

\begin{figure}
\begin{center}
\includegraphics[width=0.6\textwidth]{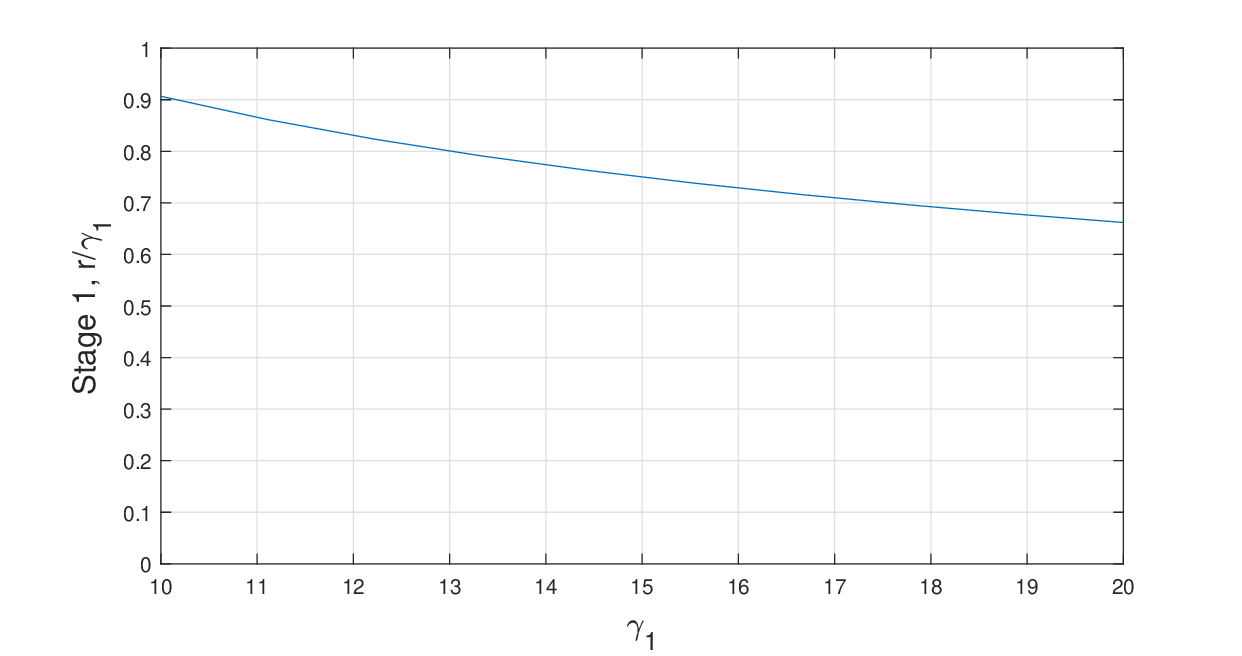}
\end{center}
\caption{Density ratio for Stage 1, normalized to $\gamma_1$.} \label{fig:rS1}
\end{figure}

Eq. (\ref{eq:r}) then gives the density ratio for Stage 1,
\begin{equation}\label{eq:rS1}
rS1 \equiv \frac{\rho_2}{\rho_1 } = \frac{\gamma_1\beta_1}{\gamma_2\beta_2} \sim \frac{\gamma_1}{\gamma_2\beta_2},
\end{equation}
We plot on Figure \ref{fig:rS1} the quantity $rS1/\gamma_1$ for a direct comparison with the isotropic case (see Appendix \ref{ap:RHIso}), namely,
\begin{equation}\label{eq:riso}
\left.\frac{\rho_2}{\rho_1 }\right|_{\mathrm{Iso}} = 2^{3/2}\gamma_1 \sim 2.82\gamma_1.
\end{equation}
With $\gamma_1 \gg 1$, we find the expansion,
\begin{equation}
rS1 = (2\gamma_1)^{2/3} + \mathcal{O}(1).
\end{equation}

\begin{figure}
\begin{center}
\includegraphics[width=0.6\textwidth]{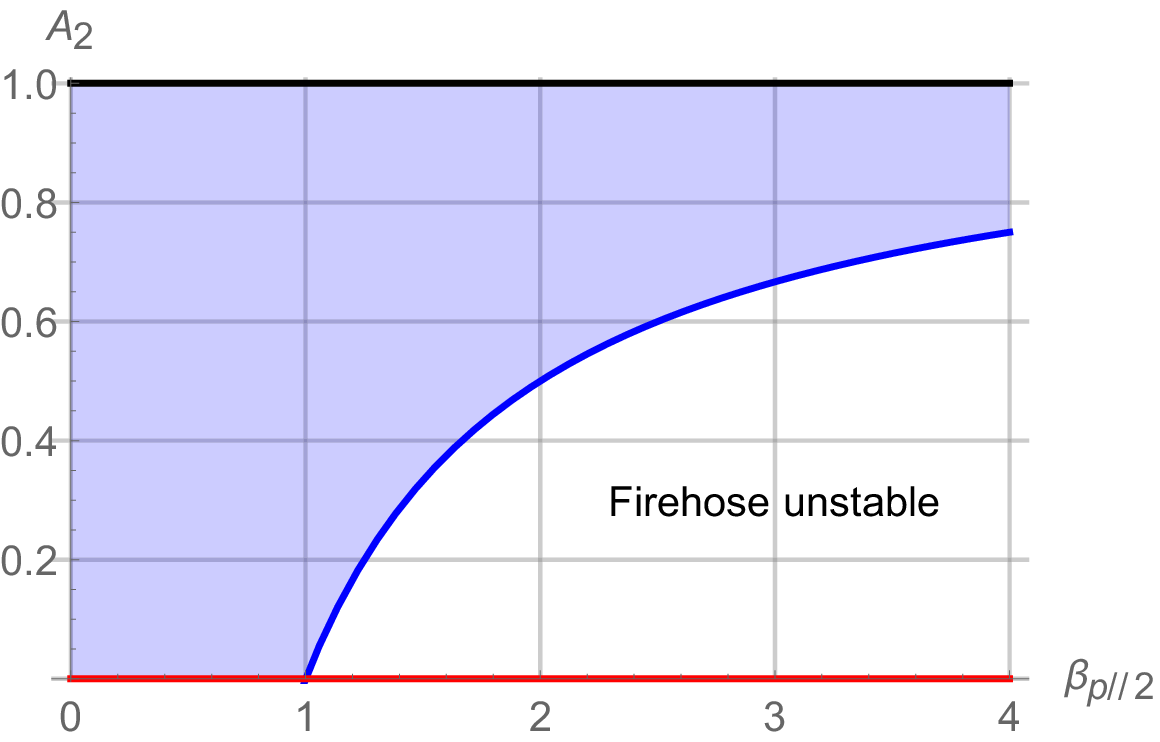}
\end{center}
\caption{Stability diagram of Stage 1. Since $T_{\perp 2} = 0$, it lies on the red line and is firehose stable within the shaded area.} \label{fig:stabS1}
\end{figure}

\subsection{Stability of Stage 1}\label{sec:stab}
Since $T_{\parallel 2} \neq 0$ while $T_{\perp 2} = 0$, the downstream anisotropy parameter in Stage 1 is
\begin{equation}
A_2 = \frac{P_{\perp 2}}{P_{\parallel 2} } = 0.
\end{equation}
As is the case for the non-relativistic system, Stage 1 can therefore be firehose unstable. The threshold of this instability is the same for the relativistic case than for the non-relativistic one \citep{Noerdlinger1968,Barnes1973}, that is,
\begin{equation}\label{eq:fire}
A_2 = 1 - \frac{1}{\beta_{p\parallel 2}},
\end{equation}
where\footnote{The notation $\beta_{p\parallel 2}$ has been chosen to clearly distinguish this magnetization parameter from the downstream parameter $\beta_2=v_2/c$.},
\begin{equation}\label{eq:beta2}
\beta_{p\parallel 2} = \frac{P_{\parallel 2}}{B_0^2/4\pi}.
\end{equation}
The stability diagram so defined is pictured on Figure \ref{fig:stabS1}. With $T_{\perp 2} = 0$, Stage 1 lies on the red line. It is firehose  stable for $\beta_{p\parallel 2} \leq 1$, that is
\begin{equation}
P_{\parallel 2} \leq \frac{B_0^2}{4\pi}.
\end{equation}
Using the dimensionless field parameter (\ref{eq:sigma}), criteria (\ref{eq:fire}) reads,
\begin{equation}\label{eq:firedimless}
\frac{\theta_{\perp 2}}{\theta_{\parallel 2}} = 1 - \gamma_1\frac{\sigma}{\theta_{\parallel 2}}.
\end{equation}
\begin{figure}
\begin{center}
\includegraphics[width=0.6\textwidth]{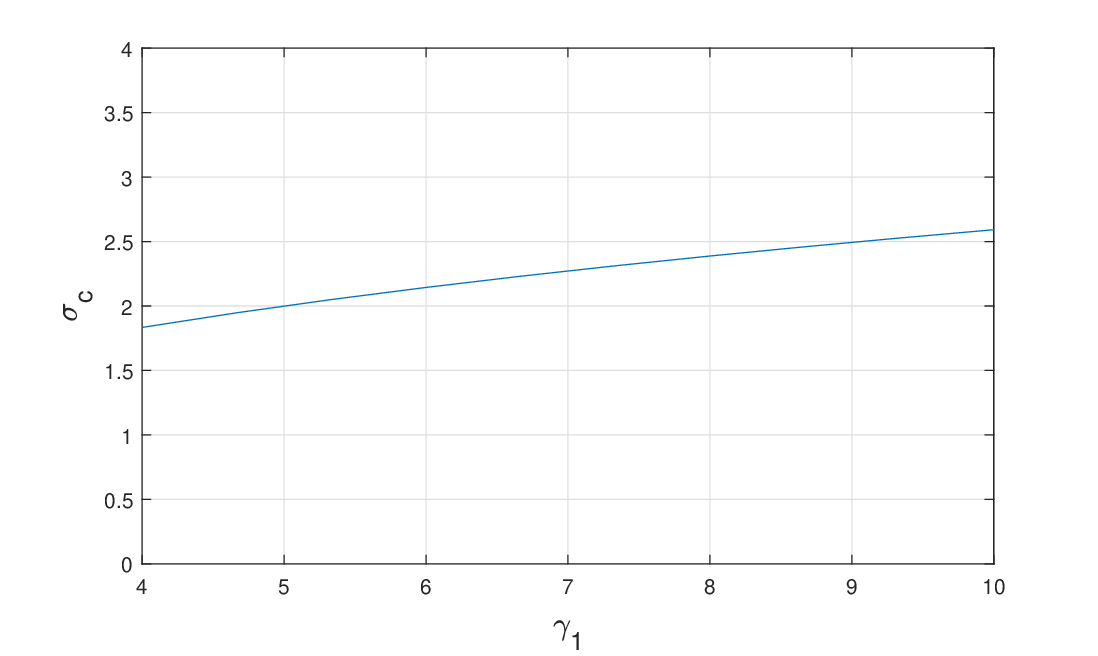}
\end{center}
\caption{Critical value $\sigma_c$ of $\sigma$ defined by Eq. (\ref{eq:SigmacS1}), above which the magnetic field stabilizes Stage 1.} \label{fig:sigmacS1}
\end{figure}
Solving right-hand-side = 0, that is,
\begin{equation}\label{eq:SigmacS1Prl}
\gamma_1\sigma_c \equiv \theta_{\parallel 2}.
\end{equation}
 allows to define $\sigma_c$, the critical $\sigma$ bellow which Stage 1 is unstable. To this extent, the value of $\theta_{\parallel 2}$ is extracted from Eq. (\ref{eq:conser2}). With $\theta_{\perp 2} =0$, and accounting for $\Gamma=4/3$ and $\beta_1 \sim 1$, it reads
\begin{equation}
 \theta_{\parallel 2} = \frac{\gamma_1}{\gamma_2}\frac{\gamma_1- \gamma_2}{2 \beta_2 \gamma_2}.
\end{equation}
 Substituting in Eq. (\ref{eq:SigmacS1Prl}) then gives,
\begin{equation}\label{eq:SigmacS1}
\sigma_c =  \frac{\gamma_1(1-\beta_2^2) - \sqrt{1-\beta_2^2}}{2 \beta_2} = \gamma_1(1-\beta_2),
\end{equation}
where $\beta_2$ is function of $\gamma_1$ via Eq. (\ref{eq:beta2S1}). $\sigma_c$ is plotted on Figure \ref{fig:sigmacS1} in terms of $\gamma_1$.

For $\gamma_1 \gg 1$, we find the expansion,
\begin{equation}
\sigma_c = (2\gamma_1)^{1/3} + \mathcal{O}(\gamma_1^{-1/3}).
\end{equation}
As the upstream kinetic energy grows, it takes an increasing field to stabilize Stage 1.

\section{Stage 2}\label{sec:S2}
We need now assess the end state of the downstream in case Stage 1 is unstable, namely, Stage 2. To do so, we come back to Eqs. (\ref{eq:conser2},\ref{eq:conser3}) and add the marginal firehose stability requirement (\ref{eq:firedimless}). The resolution follows these lines:
\begin{itemize}
  \item Use  (\ref{eq:firedimless}) to express $\theta_{\parallel 2}$ in terms of $\theta_{\perp 2}$ and $\sigma$, namely,
  \begin{equation}
  \theta_{\parallel 2} = \theta_{\perp 2} + \gamma_1 \sigma.
  \end{equation}
  \item Replace the resulting expression of $\theta_{\parallel 2}$ in (\ref{eq:conser2}) and (\ref{eq:conser3}), which gives
  \begin{eqnarray}
    \beta_2 \gamma_2^2 \left(\frac{\gamma_1}{\beta_2 \gamma_2}+2 (\gamma_1 \sigma +\theta_{\perp 2})+2 \theta_{\perp 2}\right)-\gamma_1^2 &=& 0, \\
    \beta_2^2 \gamma_2^2 \left(\frac{\gamma_1}{\beta_2 \gamma_2}+2 (\gamma_1 \sigma +\theta_{\perp 2})+2 \theta_{\perp 2}\right)-\gamma_1^2+\gamma_1 \sigma +\theta_{\perp 2} &=&  0.
  \end{eqnarray}
  \item Then eliminate $\theta_{\perp 2}$ between these two equations.
\end{itemize}
For $\gamma=4/3$ and $\gamma_1 \gg 1$, we eventually obtain an equation for $\beta_2$ only,
\begin{equation}\label{eq:beta2S2}
\sqrt{1-\beta_2^2} + \gamma_1(4\beta_2 - 3\beta_2^2 - 1) = 2  \beta_2\sigma.
\end{equation}

\begin{figure}
\begin{center}
\includegraphics[width=\textwidth]{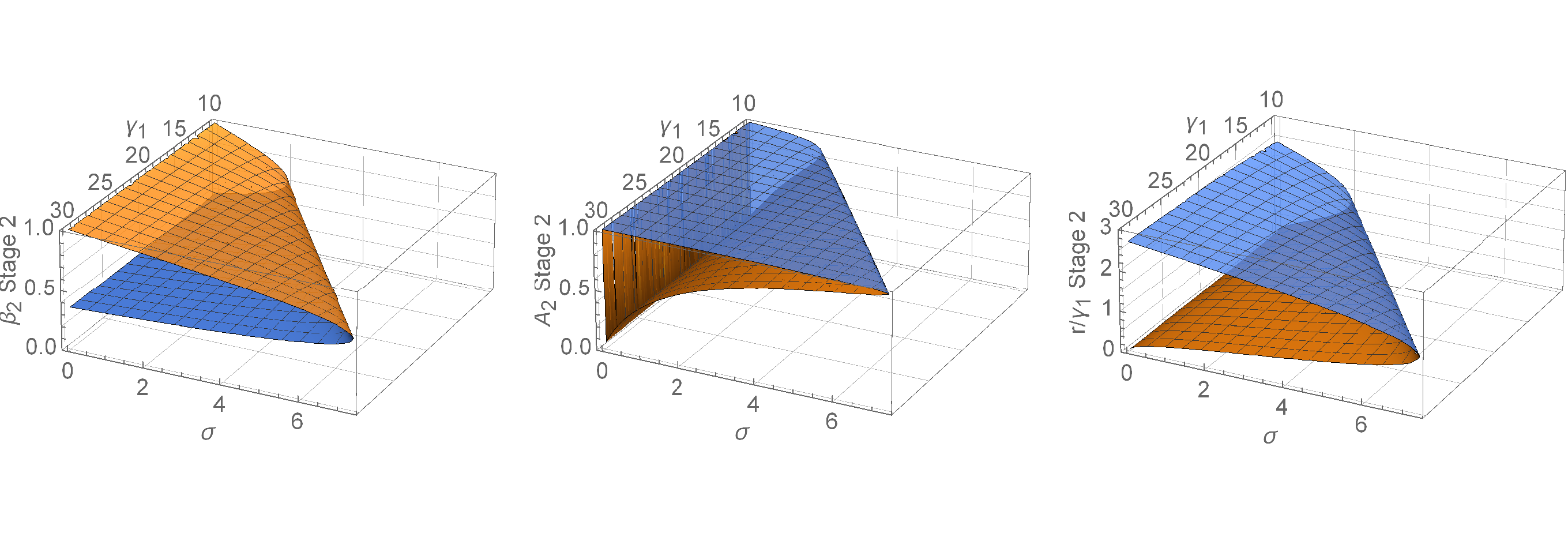}
\end{center}
\caption{Solutions of Eqs. (\ref{eq:conser2},\ref{eq:conser3},\ref{eq:firedimless}) for $\beta_2$, $A_2$ and $r/\gamma_1$ in Stage 2. The blue branch is the physical one as its $r/\gamma_1$ merges with the fluid result for $\sigma=0$ (i.e. $r/\gamma_1=2^{3/2}\sim 2.82$). The lower branch on the rightmost plot, the orange one, starts from $r<1$.} \label{fig:beta2S2}
\end{figure}

\begin{figure}
\begin{center}
\includegraphics[width=0.5\textwidth]{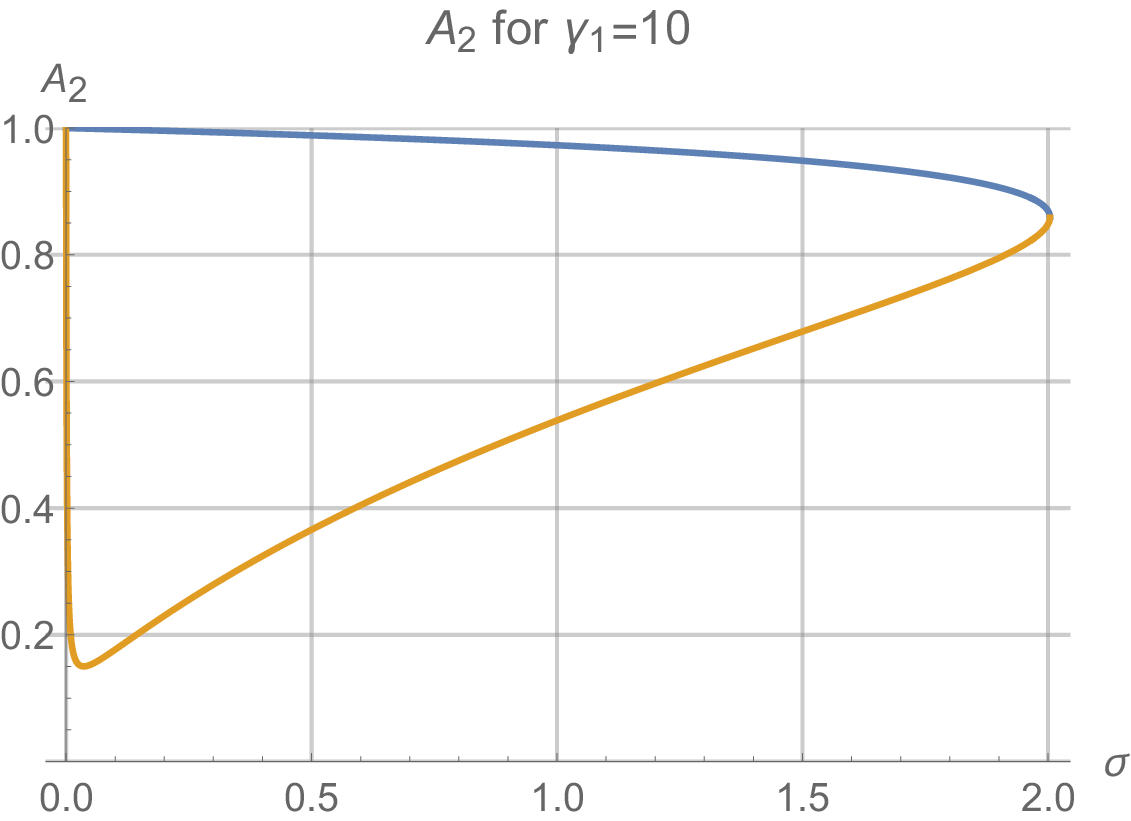}
\end{center}
\caption{Cut of Figure \ref{fig:beta2S2}-center for $\gamma_1=10$.} \label{fig:A2g1_10}
\end{figure}

The downstream anisotropy is then computed as,
\begin{equation}\label{eq:A2OK}
A_2=\gamma_1
\frac{-\beta_2^3+\beta_2+\beta_2^2 \Omega (\gamma_1+\sigma )+\Omega (\sigma -\gamma_1)}
{\Omega \sigma  \left(\beta_2^2 (\gamma_1-3)+\gamma_1-1\right)-\left(\beta_2^2-1\right) \gamma_1 (\beta_2-\Omega \gamma_1)},
\end{equation}
with $\Omega = \sqrt{1-\beta_2^2}$. It is easily checked that $A_2=1$ for $\sigma=0$.

Figure \ref{fig:beta2S2} displays the $\beta_2$ solutions of (\ref{eq:beta2S2}), $A_2$ given by (\ref{eq:A2OK}), and $r/\gamma_1$ given by (\ref{eq:rbase}), in terms of $(\sigma,\gamma_1)$. Because Eq. (\ref{eq:beta2S2}) gives 2 branches for $\beta_2$, $A_2$ and $r/\gamma_1$ display 2 branches as well. The rightmost plot show that only the upper branch, the blue one, merges with the fluid result for $\sigma=0$ (i.e. $r/\gamma_1=2^{3/2}\sim 2.82$). In contrast, the lower branch, the orange one, starts from $r<1$, a nonphysical result. Therefore, the branch we need  to consider is the blue one on the rightmost plot, that is, the lower one on the $\beta_2$ plot (leftmost plot).

Figure \ref{fig:A2g1_10} shows a cut of Figure \ref{fig:beta2S2}-center for $\gamma_1=10$. While it obviously goes to unity for $\sigma=0$, in never goes to 0 on any branch. Indeed, it hardly goes down to 0.8 with the blue branch, namely, the physical one. In other words, our model does not allow for Stage 2 solutions down to $A_2=0$, where it would merge with Stage 1.

Noteworthily,  the non-relativistic result of \cite{BretJPP2018} displayed a continuous transition between Stages 1 and 2, but only for a strong sonic shock\footnote{See Fig. 5 in \cite{BretJPP2018}.}.

\begin{figure}
\begin{center}
\includegraphics[width=0.7\textwidth]{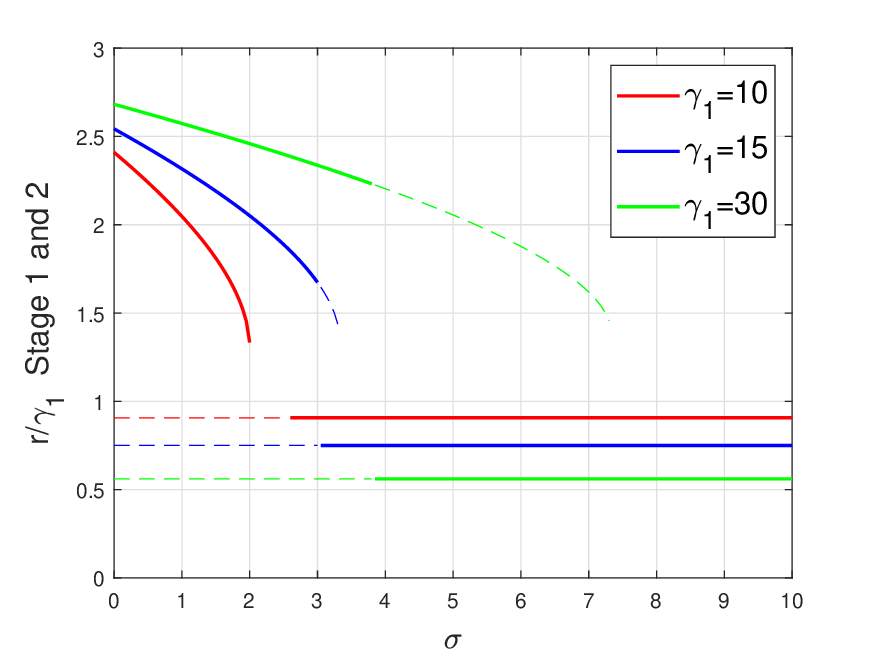}
\end{center}
\caption{Function $r(\sigma)/\gamma_1$ defined by Eq. (\ref{eq:rfinal}) for 3 values of $\gamma_1$. The horizonal lines pertain to Stage 1, where the density ratio does not depend on $\sigma$. The dashed lines picture the Stage 1 or 2 solutions which are not relevant because Stage 1 is stable, or not.} \label{fig:rS1S2}
\end{figure}

\section{Putting Stage 1 \& 2 together}\label{sec:S1S2}
We can now  put Stages 1 \& 2 together, to offer a complete solution, \emph{in the front frame}, for the density jump and any $(\gamma_1 \gg 1, \sigma)$.

Because in our scenario the plasma goes first to Stage 1, Stage 1 is the end state of the downstream as long as it is stable. The function $r(\sigma)$ is therefore defined by Stage 1 for $\sigma > \sigma_c$. If $\sigma < \sigma_c$, then the downstream settles in Stage 2, which therefore defines the density ratio. Hence,
\begin{equation}\label{eq:rfinal}
r(\sigma) = \left\{
\begin{array}{cc}
\mathrm{Stage~2}, & \sigma < \sigma_c, ~~\mathrm{Fig}.~\ref{fig:beta2S2}, \\
\mathrm{Stage ~1}, & \sigma > \sigma_c,~~\mathrm{Fig}.~\ref{fig:rS1}.
\end{array}
 \right.
\end{equation}

Figure \ref{fig:rS1S2} displays the result. We plotted the function $r(\sigma)$ defined above for 3 values of $\gamma_1$. The horizonal lines pertain to Stage 1, where the density ratio does not depend on $\sigma$. The dashed lines picture the Stage 1 or 2 solutions which are not relevant because Stage 1 is stable, or not. For $\gamma_1=15$ and 30, Stage 2 picks up (in terms of $\sigma$), when Stage 1 becomes unstable.  For $\gamma_1=10$, Stage 2 does not offer solutions up to $\sigma_c$, which is why there is a $\sigma$-gap without solution for $\sigma \in [2,2.6]$. Such has been also found to be the case for some oblique orientations of the field in the non-relativistic regime \citep{BretJPP2022b,BretJPP2024}

\section{Switching to the downstream frame and comparison with 2D3V PICs}\label{sec:PIC}
The PIC simulations of \cite{BretJPP2017} were performed using the code TRISTAN-MP \citep{Spitkovsky2005}. It is a parallel version of the code TRISTAN \citep{Buneman1993} optimized to study collisionless shocks. The spatial computational domain was 2D while the velocities and fields were rendered in 3D. The simulations implemented the ``mirror’’ method, sending a cold ($k_BT \ll mc^2$) relativistic pair plasma towards a reflecting wall. Since the shock is formed from the reflected part of the plasma interacting with the incoming part, the simulation was eventually performed in the shock downstream.

The simulation box was rectangular in the $x,y$ plane, with periodic boundary conditions in the $y$ direction and $x$ the direction of the flow. Each simulation cell was initialized with 16 electrons and 16 positrons. The time step of the simulations was $\Delta t = 0.045 \omega_p^{-1}$, with $\omega_p^2=4\pi n_0 q^2/\gamma_{1,df} m$,  where $n_0$ is the incoming plasma density in its own frame, and $m,q$  the electron mass and charge respectively.  The plasma skin depth $c/\omega_p$ was resolved with 10 simulation cells. The computational domain was $\sim 102 c/\omega_p$ in the $y$ direction, and $3600 c/\omega_p $ in the $x$ direction, that is, about $36000$ cells long. Simulations typically extended up to $t=3600 \omega_p^{-1}$.

In Eqs. (\ref{eq:conser}), all fluid quantities are defined in their own rest frame. In turn, the Lorentz and $\beta$ factors are measured in the front frame. Yet, the 2D3V PIC simulations of \cite{BretJPP2017} were performed in the downstream frame. Therefore, we need to boost Fig. \ref{fig:rS1S2} to the downstream frame for a comparison with simulations.

A key quantity in this respect is $\gamma_{1,df}$, the Lorentz factor of the upstream seen from the downstream. With
\begin{equation}
\beta_{1,df} = \frac{\beta_1-\beta_2}{1-\beta_1\beta_2},
\end{equation}
it reads
\begin{equation}\label{eq:gamma2df}
\gamma_{1,df} = \frac{1}{\sqrt{1-\beta_{1,df}^2}}= \gamma_1\gamma_2(1-\beta_1\beta_2),
\end{equation}
where the subscript $df$ stands for ``\emph{d}ownstream \emph{f}rame''.

Boosting Fig. \ref{fig:rS1S2} to the downstream frame implies boosting the density ratio, the $\sigma$ parameter, and the $\gamma_1$ parameter.

\begin{figure}
\begin{center}
\includegraphics[width=0.5\textwidth]{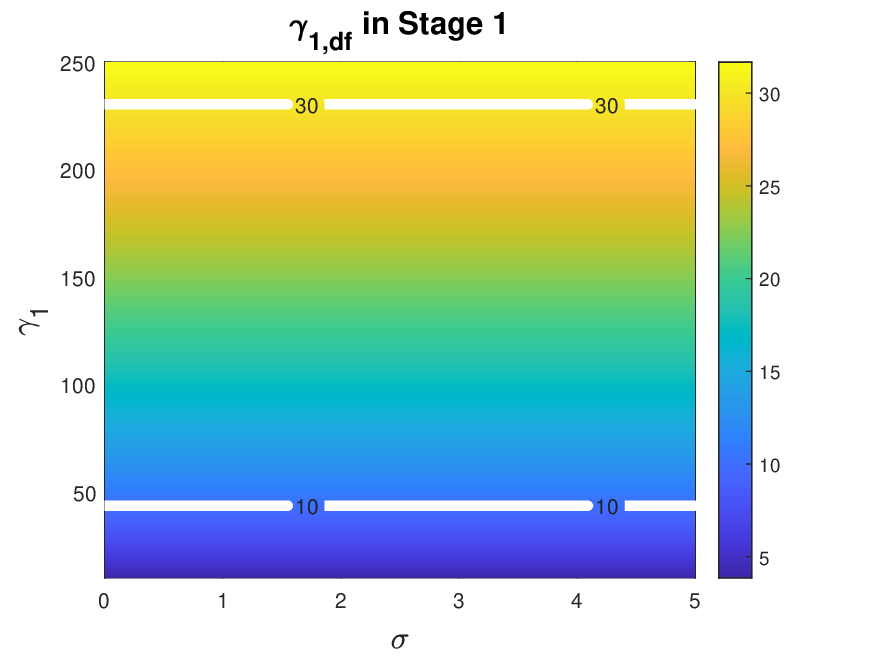}\includegraphics[width=0.5\textwidth]{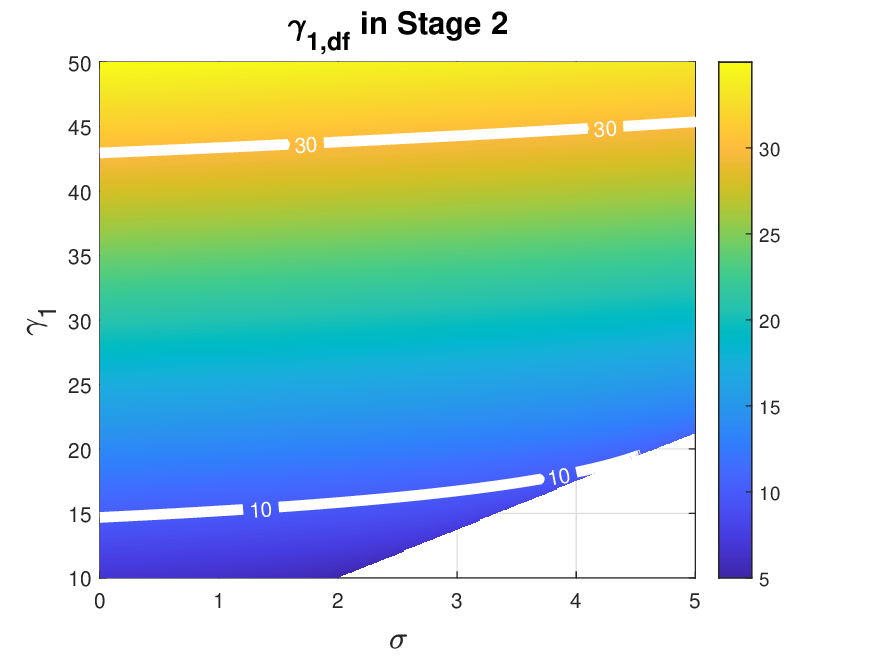}
\end{center}
\caption{Value of $\gamma_{1,df}$ in terms of $(\sigma, \gamma_1)$, for Stages 1 \& 2, with the contours of constant $\gamma_{1,df}=10$ and 30.} \label{fig:gamma12_S1S2}
\end{figure}

\begin{itemize}
  \item Regarding the density ratio, we have,
\begin{equation}
r_{df} = \frac{\rho_{2, df}}{\rho_{1, df}}.
\end{equation}
Now, $\rho_{2, df}$ is simply $\rho_2$ since it is the downstream density in its own rest frame. Regarding $\rho_{1, df}$, it reads $\rho_{2, df}= \gamma_{1,df}\rho_1$. Therefore, and still for $\gamma_1 \gg 1$,
\begin{equation}\label{eq:rdf}
r_{df} = \frac{\rho_{2, df}}{\rho_{1, df}} = \frac{\rho_2}{\gamma_{1,df}\rho_1} = \frac{1}{\gamma_{1,df}}\frac{\gamma_1\beta_1}{\gamma_2\beta_2} \sim 1+\frac{1}{\beta_2}.
\end{equation}
Hence,
\begin{equation}
r_{df} =
 \left\{
\begin{array}{cc}
1 + 1/\beta_{2,S2} & \sigma < \sigma_c, \\
1 + 1/\beta_{2,S1} & \sigma > \sigma_c.
\end{array}
 \right.
\end{equation}
where $\beta_{2, S1,S2}$ is $\beta_2$ in Stage 1 and 2, given by Eqs. (\ref{eq:beta2S1}) \& (\ref{eq:beta2S2})  respectively.
  \item The $\sigma$ parameter also needs a boost. In Eq. (\ref{eq:sigma}), the field remains unchanged since it is parallel to the boost. Only $\rho_1$ and $\gamma_1$ are modified. With $\rho_{1,df}=\gamma_{1,df}\rho_1$, we find
\begin{equation}\label{eq:sigmaboost}
  \sigma_{df} = \frac{B_0^2/4\pi}{\gamma_{1,df}~\rho_{1,df}c^2}
   = \frac{\sigma}{\gamma_1\gamma_2^2(1-\beta_1\beta_2)^2} \sim \frac{\sigma}{\gamma_1\gamma_2^2(1-\beta_2)^2} = \frac{1+\beta_2}{1-\beta_2}\frac{\sigma}{\gamma_1}.
\end{equation}
In this $\sigma$ re-scaling, $\beta_2$ is given by its Stage 1 expression when considering Stage 1, and by its Stage 2 expression when considering Stage 2.

With respect to the critical $\sigma_c$ of $\sigma$ below which Stage 1 is unstable, it reads, according to Eq. (\ref{eq:SigmacS1}), $\sigma_c  = \gamma_1(1-\beta_2)$. Using Eq. (\ref{eq:sigmaboost}) we obtain it value in the downstream frame,
\begin{equation}\label{eq:sigmadf}
  \sigma_{c, df} = 1 + \beta_2,
\end{equation}
where $\beta_2$ must be given by its Stage 1 expression since it defines the stability threshold for Stage 1. Therefore, the transition from Stage 1 to Stage 2, seen from the downstream frame, always occurs for $\sigma_{c, df} \in [1,2]$.
  \item Finally, the PIC simulations presented in \cite{BretJPP2017} considered 2 values for the upstream Lorentz factors, 10 and 30, defined \emph{in the downstream frame}. In other words, simulations varying $\sigma_{df}$ were performed at $\gamma_{1,df}$ constant. It means that to match a PIC ran at $\gamma_{1,\mathrm{PIC}}$, we need to run the calculations above for $\sigma$ and  $\gamma_1$ fulfilling,
\begin{equation}
\gamma_{1,df} = \gamma_1\gamma_2(1-\beta_2) = \gamma_{1,\mathrm{PIC}},
\end{equation}
where we used $\beta_1 \sim 1$. Figure \ref{fig:gamma12_S1S2} shows the value of $\gamma_{1,df}$ in terms of $(\sigma, \gamma_1)$, for Stages 1 \& 2. Superimposed are the contours of constant $\gamma_{1,df}=10$ and $30$. They are flat for Stage 1, where $\gamma_{1,df}$ is independent of $\sigma$, and slightly bent upward for Stage 2. As a result, we find the following correspondence,

In Stage 1,
  \begin{eqnarray}\label{eq:gamma2df_s1eq}
     \gamma_{1,df} = 10 &\Rightarrow& \gamma_1 = 43, \nonumber\\
     \gamma_{1,df} = 30 &\Rightarrow&  \gamma_1 = 230.
   \end{eqnarray}

In Stage 2,
    \begin{eqnarray}
     \gamma_{1,df} = 10 &\Rightarrow& \gamma_1 \sim 16, \nonumber\\
     \gamma_{1,df} = 30 &\Rightarrow&  \gamma_1 \sim 43.
   \end{eqnarray}
It has been checked that for Stage 2, considering a constant $\gamma_1$ hardly affects the result.
\end{itemize}

With Eq. (\ref{eq:gamma2df}) reading $\gamma_{1,df} = \gamma_1\gamma_2(1-\beta_1\beta_2)$, one could expect in Eq. (\ref{eq:gamma2df_s1eq}) for Stage 1, a value of $\gamma_1$ lower than 230. Indeed, \emph{if}, as hinted in the Appendix, $\beta_2 \sim 1/3$ and $\gamma_2 \sim 1$, then with $\gamma_1=230$, $\gamma_{1,df} \sim 230 \times 1 \times 2/3 = 153 > 30$. But in the present case, the value of $\beta_2$ has to be the one derived in Section \ref{sec:S1}, for the \emph{anisotropic} downstream in Stage 1, not the \emph{isotropic} case treated in the Appendix. And for $\gamma_1 = 230$, Stage 1 has $\beta_2 \sim 0.967$, which explains how the product $\gamma_1\gamma_2(1-\beta_1\beta_2)$ is reduced to just 30.

\begin{figure}
\begin{center}
\includegraphics[width=0.6\textwidth]{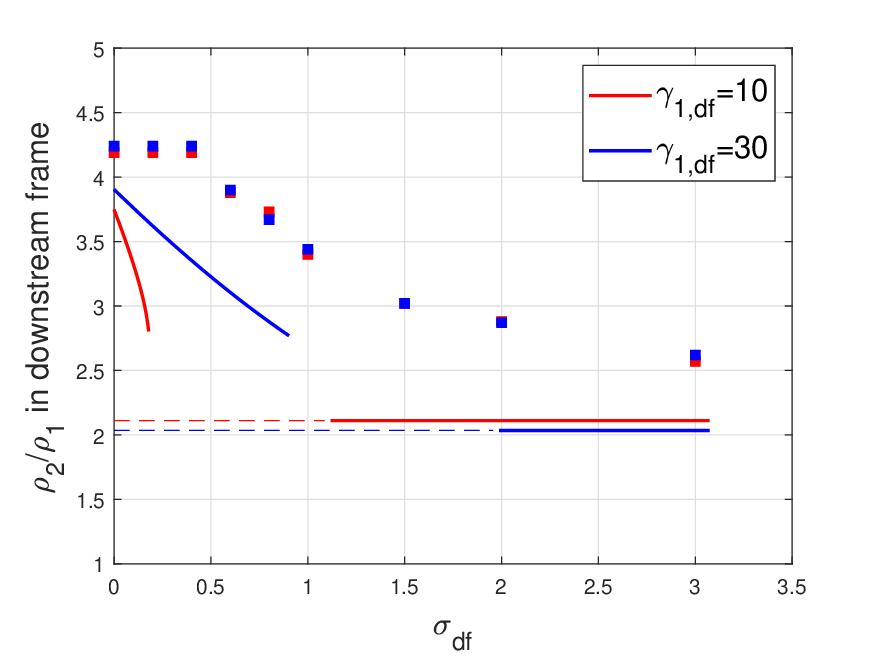}
\end{center}
\caption{Density ratio measured in the downstream frame, $r_{df}$, in terms of $\sigma_{df}$. The transition between the 2 stages occurs for the critical $\sigma_{c, df}$ given by Eq. (\ref{eq:sigmadf}). The squares show the results of the PIC simulations performed in \cite{BretJPP2017}.} \label{fig:PIC}
\end{figure}

Figure \ref{fig:PIC} is the end result of this article. It displays the density ratio measured in the downstream frame, $r_{df}$, in terms of $\sigma_{df}$. The transition between the 2 stages occurs for the critical $\sigma_{c, df}$ given by Eq. (\ref{eq:sigmadf}). The squares shows the results of the PIC simulations performed in \cite{BretJPP2017}\footnote{See Figs. 3 \& 6 in \cite{BretJPP2017}.}.

Note that Stage 1 always defines the density ratio for large fields $\sigma_{df} > \sigma_{c, df}$. It is therefore instructive to derive an expansion of (\ref{eq:rdf}) in Stage 1. We find,
\begin{eqnarray}\label{eq:rlim}
r_{df, S1} = 1 + \frac{1}{\beta_{2,S1}} &=& 2 + \left( \frac{\sqrt{2}}{\gamma_1} \right)^{2/3} + \mathcal{O}(\gamma_1^{-4/3}) \nonumber \\
&=& 2 + \frac{1}{\gamma_{1, \mathrm{PIC}}} + \mathcal{O}(\gamma_{1, \mathrm{PIC}}^{-2}) .
\end{eqnarray}

\section{Conclusion}
In this work, we have implemented the relativistic version of the model presented in \cite{BretJPP2018}. The system considered is a collisionless shock in pair plasma, with a parallel magnetic field. Since numerical simulations have been carried out in \cite{BretJPP2017}, we are able to compare the results of the present theory with these simulations.

After adapting to the relativistic case the theory presented in \cite{BretJPP2018}, and boosting the results to the downstream frame, we reach,  on Figure \ref{fig:PIC}, the comparison of the relativistic theory with the PIC simulations.

The  agreement is satisfactory. Our model reproduces the fall of the density ratio over the correct the $\sigma_{df}$ range, with an asymptotic value close to 2 in the strong field regime $\sigma_{df} > \sigma_{c, df}$.

We observe 2 kinds of discrepancies on Fig. \ref{fig:PIC}, between the theory and  the simulations. \emph{(a)} There is a systematic underestimation  of the density ratio of the theory with respect to simulations and \emph{(b)}, the PIC simulations find solutions where the model does not offer any, like for $\sigma_{df} = 1.5$ with $\gamma_{1,df}=30$. Let us briefly comment on each point.
\begin{enumerate}
  \item In the absence of a magnetic field, the density jump measured  in the downstream is given by  (see Appendix \ref{ap:RHIso}),
  \begin{equation}\label{eq:JumpRamesh}
  r_{df} = \frac{\Gamma}{\Gamma-1}.
  \end{equation}

  For an adiabatic index $\Gamma= 4/3$, this gives $r_{df}=4$. Yet, as seen on Fig. \ref{fig:PIC},  PIC simulations for weak field give a slightly higher value, $r_{df}\sim 4.3$. Such an overshoot could be interpreted as the PIC effectively simulating an adiabatic index $\Gamma=1.3$, slightly lower than the fully relativistic $\Gamma=4/3 \sim 1.33$ considered in the theory\footnote{Setting $\Gamma = 1.3$ in Eq. (\ref{eq:JumpRamesh}) yields $r_{df}=4.3$.}. Yet, this solution seems improbable, for if the downstream were not fully relativistic in the simulation, its adiabatic index would rather lean towards an \emph{higher} value, namely  $\Gamma=5/3 \sim 1.66$.

  Why, then, is the density ratio in the PIC simulations higher than Eq. (\ref{eq:JumpRamesh}) predicts? The most probable answer is related to accelerated particles. It has been known for long that collisionless shocks can accelerate particles \citep{Blandford78}. And it has been equally known for long than their effect on the shock is to raise its density ratio \citep{Berezhko1999,Stockem2012}. To date, this is the only known mechanism capable of raising the density ratio above its MHD value (see \cite{BretApJ2020} and references therein). Therefore, strictly speaking, the PICs presented here are \emph{not} a simulation of the theory, since the theory ignores particles acceleration, which simulations necessarily include. The PIC simulations eventually simulate the system discussed here, embedded in a bath of cosmic rays.
  \item Besides ignoring particle acceleration, other differences between the PIC simulations and the theory are noticeable. For example, Figs. 3(b), 4(b) \& 6(b) of \cite{BretJPP2017} show that while the upstream temperature is nearly 0, the downstream is definitely anisotropic in the PICs but its perpendicular temperature is never 0, even in the strong field regime. This means that Stage 1, as described here, is not fully realized. Probably due to the influence of cosmic rays,  the simulations do not stick to the present theory, even though they follow its trend. It is therefore not surprising that the simulations  find downstream solutions even when our rigid, ``cosmic rays-less'' theory, does not allow any. Future works could focus on some detailed analysis of the particle distribution in the PIC. Yet, as mentioned above, it could be challenging to run some PIC simulations cancelling the cosmic rays they necessarily produce.
\end{enumerate}

There is a striking similarity between the non-relativistic and relativistic theories. In the non-relativistic theory, the density jump in strong field tends to $r=2$. This is an important departure from the MHD prediction, since the latter predicts a field-independent density jump for a parallel shock of $r=4$ (strong sonic shock).

The same pattern is observed here. Relativistic MHD still predicts a field-independent density jump for this parallel geometry, which amounts to a density jump measured in the downstream  $r_{df} \sim 4$. In contrast, the relativistic theory presented here, and confirmed by simulations, gives a density jump also tending towards $r_{df}=2$.

It is likely that, as in the non-relativistic case \citep{BretJPP2022b,BretJPP2024}, the departure from MHD is maximum for this obliquity of the magnetic field. This could be confirmed in future works.

Extending this work to finite upstream pressures would require modifying the left-hand-side of Eqs. (\ref{eq:conserv2},\ref{eq:conserv3}), before implementing similar calculations following the same road map. While this could be the goal of some future work, the result would probably be a lowering of the density ratio with diminishing sonic Mach number, as was observed in the non-relativistic case \citep{BretJPP2018}.

\section*{Funding}
A.B. acknowledges support by the Ministerio de Econom\'{i}a  y Competitividad of Spain (Grant No. PID2021-125550OB-I00).
R.N. was partially supported by the Black Hole Initiative at Harvard University, which is funded by the Gordon and Betty Moore Foundation grant 8273, and the John Templeton Foundation
grant 62286.

\section*{Declaration of Interests}
The authors report no conflict of interest.

\appendix
\section{Jump conditions for an isotropic fluid with $P_1=0$.}\label{ap:RHIso}
Although the simple case of a relativistic shock in an isotropic fluid has been abundantly discussed in literature \citep{Taub1948,LandauFluid,Anile2005}, we here present a simple derivation of the jump conditions, for $P_1=0$ and $\gamma_1 \gg 1$,  using the present notations \citep{Narayan2012}.

We start from Eqs. (\ref{eq:conserv1}-\ref{eq:conserv3}) where only $P_1=0$ has been assumed. Setting $\theta_{\perp 2}=\theta_{\parallel 2} = \theta_{2}$ to reflect isotropy gives,
\begin{eqnarray}
\rho_1 \gamma_1\beta_1 &=& \rho_2 \gamma_2\beta_2,  \label{eq:conservIso1} \\
  \gamma_1^2\beta_1  &=&
   \gamma_2^2\beta_2\left(\frac{\Gamma}{\Gamma - 1}\theta_2 +   \frac{\rho_2}{\rho_1} \right), \label{eq:conservIso2}\\
  \gamma_1^2\beta_1^2   &=& \gamma_2^2\beta_2^2\left(\frac{\Gamma}{\Gamma - 1}\theta_2 +   \frac{\rho_2}{\rho_1}  \right) +\theta_2. \label{eq:conservIso3}
\end{eqnarray}
Computing $(\mathrm{\ref{eq:conservIso3}}) - \beta_2 \times (\mathrm{\ref{eq:conservIso2}})$ gives,
\begin{equation}
\gamma_1^2\beta_1(\beta_1-\beta_2)=\theta_2.
\end{equation}
Using this expression for $\theta_2$ in (\ref{eq:conservIso2}) gives,
\begin{equation}
\gamma_1 - \gamma_2 = \frac{\Gamma}{\Gamma - 1}\gamma_1\gamma_2^2\beta_2(\beta_1-\beta_2).
\end{equation}
Since $\gamma_1 \gg 1$, $\beta_1 \sim 1$. In addition, because the downstream has to be subsonic while the speed of sound in a relativistic gas is $c/\sqrt{3}$, we know $\beta_2 < 1/\sqrt{3}$. Hence, $\gamma_2 \sim 1$ so that $\gamma_1 \gg \gamma_2$. The equation above therefore simplifies to,
\begin{eqnarray}
\gamma_1 - \gamma_2 &\sim & \gamma_1 \nonumber \\
                     &\sim & \frac{\Gamma}{\Gamma - 1}\gamma_1\gamma_2^2\beta_2(1-\beta_2) \nonumber \\
                     & \Rightarrow&  ~~ \beta_2 = \Gamma - 1,
\end{eqnarray}
where $\gamma_2^2 = (1-\beta_2^2)^{-1}$ and $\beta_1 \sim 1$ has been used. This expression for $\beta_2$ is identical to Eq. (16) of \cite{Kirk1999}.
Then,
\begin{equation}
\gamma_2 = \frac{1}{\sqrt{1-\beta_2^2}} = \frac{1}{\sqrt{(2-\Gamma ) \Gamma }},
\end{equation}
and from (\ref{eq:rbase}),
\begin{equation}\label{eq:rbaseiso}
r = \frac{\rho_2}{\rho_1} \sim \frac{\gamma_1}{\beta_2\gamma_2} = \gamma_1 \frac{\sqrt{\Gamma (2-\Gamma)}}{\Gamma}.
\end{equation}
Finally,
\begin{equation}
\gamma_{1,df} \sim \gamma_1(1-\beta_2) = \gamma_1\sqrt{\frac{2-\Gamma}{\Gamma}},
\end{equation}
and,
\begin{equation}\label{eq:rbasedf}
r_{df} = \frac{r}{\gamma_{1,df}} = \frac{\Gamma}{\Gamma-1}.
\end{equation}
For $\Gamma=4/3$, Eqs. (\ref{eq:rbaseiso},\ref{eq:rbasedf}) give $r = 2^{3/2}\gamma_1$ and $r_{df} = 4$ respectively. Note that PIC simulations where velocities are rendered in only 2 dimensions, give a density ratio of $r_{df} = 3$, corresponding to $\Gamma=3/2$.

In \cite{BretJPP2017}, velocities were rendered in 3D, as explained in Section \ref{sec:PIC}. If they had been rendered in 2D, the density jump would have been 3. For a direct comparison of the 2D \& 3D cases in PIC simulations, see Fig. 3 of \cite{BretPoP2014}.

\bibliographystyle{jpp}
\bibliography{BibBret}

\end{document}